\documentclass[conference]{IEEEtran}
\IEEEoverridecommandlockouts
\usepackage{cite}
\usepackage{amsmath,amssymb,amsfonts}
\usepackage{algorithmic}
\usepackage{graphicx}
\usepackage{textcomp}
\usepackage{xcolor}
\usepackage{balance}
\usepackage{makecell}
\usepackage{hyperref}
\usepackage{url} 
\usepackage{listings} 
\usepackage{float}

\usepackage[letterpaper,%
            left=0.75in,right=0.75in,top=0.75in,bottom=1in,%
            footskip=.25in,bindingoffset=0.2in]{geometry}

\def\BibTeX{{\rm B\kern-.05em{\sc i\kern-.025em b}\kern-.08em
    T\kern-.1667em\lower.7ex\hbox{E}\kern-.125emX}}

\begin{document}

\makeatletter
\newcommand{\linebreakand}{%
 \end{@IEEEauthorhalign}
 \hfill\mbox{}\par
 \mbox{}\hfill\begin{@IEEEauthorhalign}
}
\makeatother

\title{Data Extraction, Transformation, and Loading Process Automation for Algorithmic Trading Machine Learning Modelling and Performance Optimization

\thanks{We would like to acknowledge and thank the Post Degree Diploma program in Data Analytics and the Work on Campus programs at Langara College and the Computer Science Department at Okanagan College for supporting our research.}
} 
\author{

    \IEEEauthorblockN{Nassi Ebadifard}
    \IEEEauthorblockA{\textit{Computer Science} \\
   \textit{Okanagan College}\\
    Kelowna, Canada \\
   0009-0002-9087-5259}

   \and
   
\IEEEauthorblockN{Ajitesh Parihar}
    \IEEEauthorblockA{\textit{Computer Science} \\
   \textit{Okanagan College}\\
    Kelowna, Canada \\
   0009-0001-3162-1470}

   \and
   
     \IEEEauthorblockN{Youry Khmelevsky}
    \IEEEauthorblockA{\textit{Computer Science} \\
   \textit{Okanagan College}\\
    Kelowna, Canada \\
   0000-0002-6837-3490}
    
    \linebreakand

    \and
    
\IEEEauthorblockN{Ga\'etan Hains}
   \IEEEauthorblockA{ \textit{LACL} \\
   \textit{Université Paris-Est}\\
   Créteil, France \\
   0000-0002-1687-8091} 

\and
    
\IEEEauthorblockN{Albert Wong}   \IEEEauthorblockA{\textit{Mathematics and Statistics} 
  \\   \textit{Langara College}\\
   Vancouver, Canada \\
   0000-0002-0669-4352}

\and 
    \IEEEauthorblockN{Frank Zhang}
    \IEEEauthorblockA{\textit{School of Computing} \\
   \textit{University of the Fraser Valley}\\
   Abbotsford, Canada \\
    0000-0001-7570-9805}
      
}

\maketitle

\begin{abstract}
A data warehouse efficiently prepares data for effective and fast data analysis and modelling using machine learning algorithms. This paper discusses existing solutions for the Data Extraction, Transformation, and Loading (ETL) process and automation for algorithmic trading algorithms.  Integrating the Data Warehouses and, in the future, the Data Lakes with the Machine Learning Algorithms gives enormous opportunities in research when performance and data processing time become critical non-functional requirements. 
 
\end{abstract}


\section{Introduction}


The popularity of machine learning models and algorithms has been increasing greatly over the past decade. They will continue to be used even more, especially with the tremendous success of Large Language Models (LLM) in data analysis \cite{makridakis2023large}. Professionals like researchers and analysts incorporated machine learning into daily lives. From corporations to individuals, the use of machine learning is applicable in a wide range of settings \cite{wong2023short}.

To improve the performance of the machine learning algorithms for stock price forecasting, we decided to implement two layers of the databases: (1) an OLTP DBMS system for the data collection and (2) following data migration and transformation into a data warehouse (DW) employing an efficient ETL process.

The advancement of automation in the ETL process aims to reduce human errors and streamline the process. This is usually done using tools and software to automate data extraction, cleansing, and transformation tasks. Automation can also schedule and monitor ETL jobs, ensuring they run on time and produce accurate results.

First, this paper analyzes existing works related to Algorithmic Trading data collection and the following transformation into a DW. Then, we describe our prototype design and development for automatic data collection, which involves using various tools and technologies. The main contributions of this paper are (1) automation of data collection in an OLTP database \textbf{in a cloud }and (2) \textbf{automated} ETL process for the data collection, transformation and loading in a DBMS and then transfer into DW (3) ML performance improvement in the domain of Algorithmic Trading Machine Learning Modelling. The following steps will relate to the data pre-aggregation, drastically improving ML performance (see more information in our previous works here \cite{Dhanjal2004a, Govorov2005, 
Khmelevsky2009swp, Khmelevsky2009gis, Khmelevsky2004_9iAS, Khmelevsky2010d, Khmelevsky2011res, Khmelevsky2011distance, Khmelevsky2011international, Khmelevsky2012automatic, 
Khmelevsky2013strategies, Khmelevsky2014Minecraft, khmelevsky2015hybrid, Khmelevsky2015Minecraft, Khmelevsky2015bot, Khmelevsky2016TenYears, Khmelevsky2016paradigm, Khmelevsky2016reporting, Khmelevsky2016sport, Khmelevsky2016game, Khmelevsky2016GPN, Khmelevsky2016Anew, Khmelevsky2016biometric, Khmelevsky2017Astochastic, Khmelevsky2017utilizing, 
Khmelevsky2017gaming, Khmelevsky2017b, Khmelevsky2019state, Khmelevsky2023DW}).  

For the 1st testing prototype in the Cloud, we used Oracle APEX to test data collection and cleaning. The 2nd system's prototype was built using Compute Canada research resources (now this is The Digital Research Alliance of Canada\cite{alliancecan.ca}). We used a Python Programming language and a PostgreSQL OLTP database system in the Cloud to gather publicly available data from commercial companies to test the effectiveness of ML models. But we found that the programming environment was a little bit complicated. The 3rd system's prototype is built using PHP code, hosted on GitHub\cite{github.com} for source code version control and user documentation storage. The data collection process is being executed on commercial web hosting at Ionos.com (Ionos)\cite{ionosPlans}, and the data is stored in a MySQL database for prototyping purposes only (no other DBMS systems are available now for web hosting purposes). We used Yahoo Finance (Yahoo Finance API has been discontinued since 2017) and Financial Modelling Prep \cite{financialModelingPrep}. After the prototyping is completed, we plan to move our project to PostgreSQL OLTP \& DW in a virtual server environment with a data storage size of up to 12 TB.  We are still using an ``Expert" Web-hosting plan at Ionos.  Internal documentation and extracted data are stored on HiDrive\cite{hiDrive}. The data collection is executed daily via cron jobs from two different sources of information. Finally, the data is processed and modelled on a personal computer, but the next step is to use available Cloud Computing and multiple GPU processing resources.


\section{Existing Works}


Machine learning for predicting changes in stock price has gained popularity in the last decades. However, there has been a concerning neglect of the efficiency of algorithmic trading. \cite{Dubey2022}. Data is an essential component of creating a successful model for algorithmic trading. However, some issues arise and give challenges \cite{Martinez2022}. According to Théate and  Damien \cite{theate2021application}, training the resulting reinforcement learning (RL) agent is completely based on ``generating artificial trajectories from a limited stock market historical data set".

According to Haryono et al. \cite{haryono2020comparison}, ETL (Extract, Transform, Load) and ELT (Extract, Load, Transform) are the two primary data processing methods for implementing a data warehouse. Choosing the correct method is difficult because consideration of a company's cost, efficiency, and procedure plays a vital role in determining the implementation of a DW. The data collected for training the models would be collected from different sources and would be heterogeneous. Yulianto concludes in their research \cite{yulianto2019extract} that the heterogeneity of data from different sources can be dealt with by designing the ETL Process to include this ETL Data Flow: ``Extract, Clean, Conform, and Delivery/Load". The ETL Process contains various steps to address the issues of heterogeneous data. It is a common issue where institutions must capture heterogeneous data from different sources to store it in single or multiple data stores. In addition, different IT systems that make use of distinct data structures also pose a big challenge. Azeroual et al.\cite{azeroual2019etl} concluded that implementing the ETL process is the solution to these challenges. Doing so provides a data filtering, aggregation, cleaning and verification method. We have chosen the object-oriented design approach for designing an efficient ETL process. Object-oriented design is a well-established and proven method for creating effective and robust systems, reinforcing our decision to use this approach.

Ali\cite{ali2018next} created an extendable ETL framework that would address the challenges that Big Data gives rise to. They proposed a user-defined-functions (UDF) Component to address the lack of support for UDFs and the optimizations they discovered in the ETL frameworks they observed.

Many ETL tools have been developed. Patel et al. compared the features of different tools in their literature review \cite{patel2020progressive}. They pointed out the development of various categories of ETL tools, some of them being:`` code-based, GUI-based, cloud-based, Metadata support, Real-time support, and batch processing".

According to Michael et al., \cite{michael2020improved}, a properly documented and well-designed ETL system is essential to create a successful Data Warehouse project. They designed an ETL system that used PowerBi with MS Excel and Visual Studio 2017 with SQL Server to execute ETL operations. However, a system like this would have some shortcomings. Such implementations would suffer from non-optimal performance and programming difficulty compared to a custom-coded approach. The efficiency of the ETL System is a high priority. According to Neepa Biswas et al., \cite{biswas2020efficient}, GUI-based products sold by vendors are the go-to for most organizations for their ETL solution. However, they also discovered that custom-coded ETL provides better performance and efficiency in specific cases. They compared the performance of four code-based ETL tools, namely: Pygrametl, Petl, Scriptella and ETL. Considering all this, we decided to use tools like PHP and MySQL Data Warehouse to accommodate a big dataset to analyze time series trajectories for our prototype.

We have also discovered that there is a lack of use of cloud resources in the papers that we reviewed, which severely limits their scalability and flexibility. We have incorporated multiple cloud services in our prototype to address these issues and improve the performance of the ETL process.



\section{ETL Process Automation Architecture and Design}

We model and design a system using php, GitHub, Ionos, MySQL and HiDrive for the ETL process.

\subsection{Project Prototype Design}
A simplified project prototype Use Case diagram is shown in Fig.~\ref{fig: Use Case}, where data analysis is performed on both OLTP and DW DBMS systems.

\begin{figure}[ht]
\centering
\includegraphics[width=0.5\textwidth]{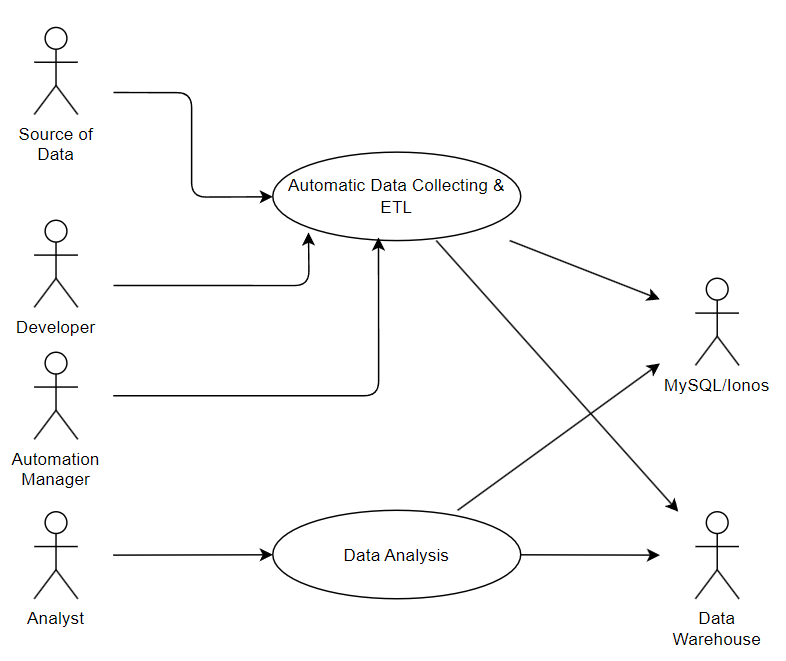}
\caption{A Simplified Project Prototype Use Case Diagram}
\label{fig: Use Case}
\end{figure}


The prototype design consists of the following components (see Fig.~\ref{fig: Prototype Design}):

\begin{figure}[ht]
\centering
\includegraphics[width=\columnwidth]{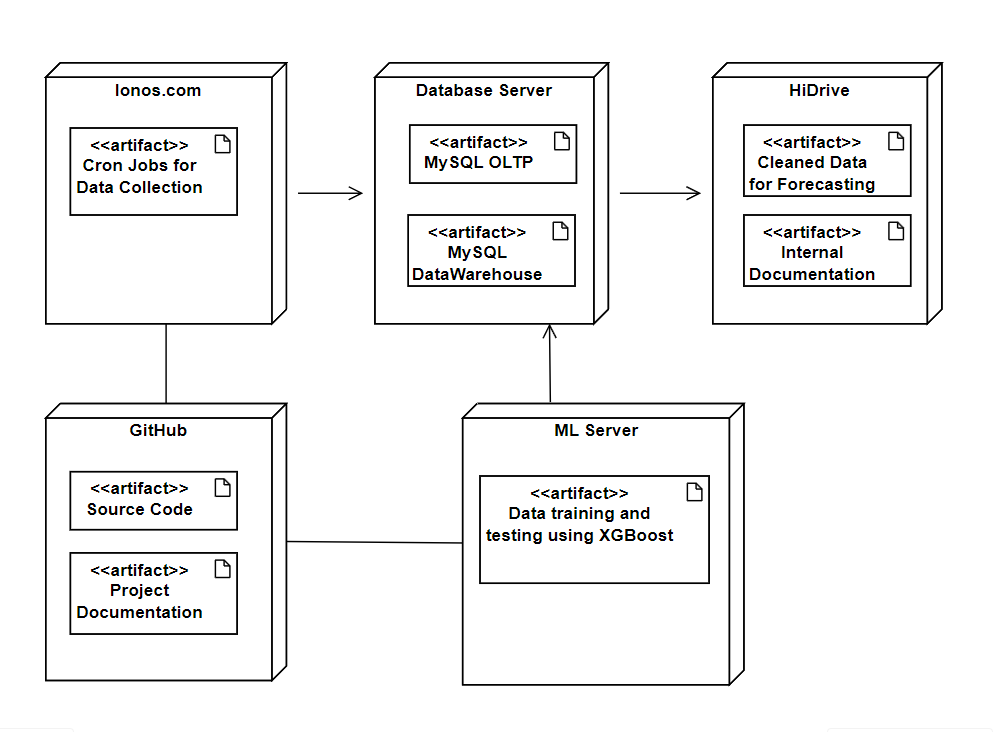}
\caption{Project Prototype Design}
\label{fig: Prototype Design}
\end{figure}

\begin{itemize}
    \item Data Collection from various public and commercial sources. 
    \item Source code and the development documentation are stored on Github\cite{githubAlgorithmicTrading}.
    \item MySQL OLTP and DW instances to store data and prepare for the forecasting using ML algorithms (see our initial forecasting results here: \cite{wong2023short}, \cite{Wong2023}).
    \item HiDrive to store extracted data and internal documentation.
    \item Server for data processing and data modelling (Cron jobs would be used in the future for data extraction and processing on a cloud server).
\end{itemize}


During the Project Prototype design and implementation, we discovered that in the context of ETL, the Data Analyst is the primary actor rather than the Developer and the Automation manager (see Fig.\ref{fig: Use Case}). The system is being designed to serve the Data Analyst's goals primarily. The Developer writes the source code, and the Automation manager maintains the Automatic data collection and ETL. The Analyst extracts Data from HiDrive and MySQL to use it.\\

The discovered actors and their roles are shown in Fig.~\ref{fig: Use Case}:
    \begin{itemize}
        \item \textbf{Developer} --- A user who creates software applications using programming languages.
        \item\textbf{Automation Manager} --- A user who sets up and manages the automated data collection and processing system.
        \item\textbf{Data Analyst} --- A user who uses the processed data for analysis.
    \end{itemize}

For the data collection and processing, we have conceived a flow for data collection from a PHP file hosted on Ionos.com. First, The Automation Manager sets up the automatic cron jobs on the Ionos and configures the
system to run the PHP file and data processing scripts automatically at specified intervals. The PHP file and data processing scripts are designed and tested for accuracy and performance. Subsequently, this data is stored within a MySQL database, also hosted on Ionos. For this purpose, the data analyst creates scripts to train and model the data, using JupyterLab, the ultimate objective of processing and modelling the data to enable comprehensive analysis. The MySQL database is designed to handle the expected amount of data and is backed up regularly. The processed data is then stored in a designated location in the HiDrive for easy access and collaboration. The HiDrive is accessible and has sufficient storage space for the documentation, code, and data. Finally, the data analyst can access the processed data and perform analysis. Central to this is the imperative for complete automation of the data collection process, encompassing the data itself, the underlying code, and associated documentation. We decided that GitHub, a central repository and version control platform, will be used for project management and collaborative development. Overall, the system is designed with security, appropriate access controls, and protection against data breaches.

We have also identified and mitigated a few expected issues. 
Suppose the PHP file or data processing scripts encounter errors during execution. In that case, the system triggers an email notification promptly dispatched to the Automation Manager, who assumes responsibility for resolving these issues. In instances where updates or modifications to the processed data become requisite, the Developer, as the key custodian of the codebase, undertakes necessary changes to the scripts and configures the system for their re-execution.
Additionally, when there is a demand for integrating new data sources into the system's framework, The Developer updates the PHP file, and the Automation Manager adjusts the database schema to accommodate the new data inputs. Moreover, a cloud-based password manager, like RoboForm\cite{roboform}, is judiciously employed as a robust safeguarding mechanism to enhance the system's security and safeguard the data. This proactive measure ensures the confidentiality and integrity of the sensitive data within the system.

The project Use Case diagram (Fig.~\ref{fig: Use Case}) demonstrates the iterative and incremental approach to software development, where the system is developed and tested in small iterations and where potential risks and issues are identified and mitigated throughout the development process. It also focuses on collaboration and version control, where the code, documentation, and data are stored in a central location for easy access and collaboration.

\section{The Automation Process Overview}
Simplified data groups for the analysis and forecasting collected from different sources are shown in Fig.~\ref{fig: Data Groups}. The base class ``Companies" contains symbol, name, sector and industry attributes. The sub-classes ``Stock" and ``Commodities" extend the ``Companies" class and add the price and volume attributes. The date class contains the capture timestamp of the ``Companies" class.

\begin{figure}[ht]
\centering
\includegraphics[width=0.45\textwidth]{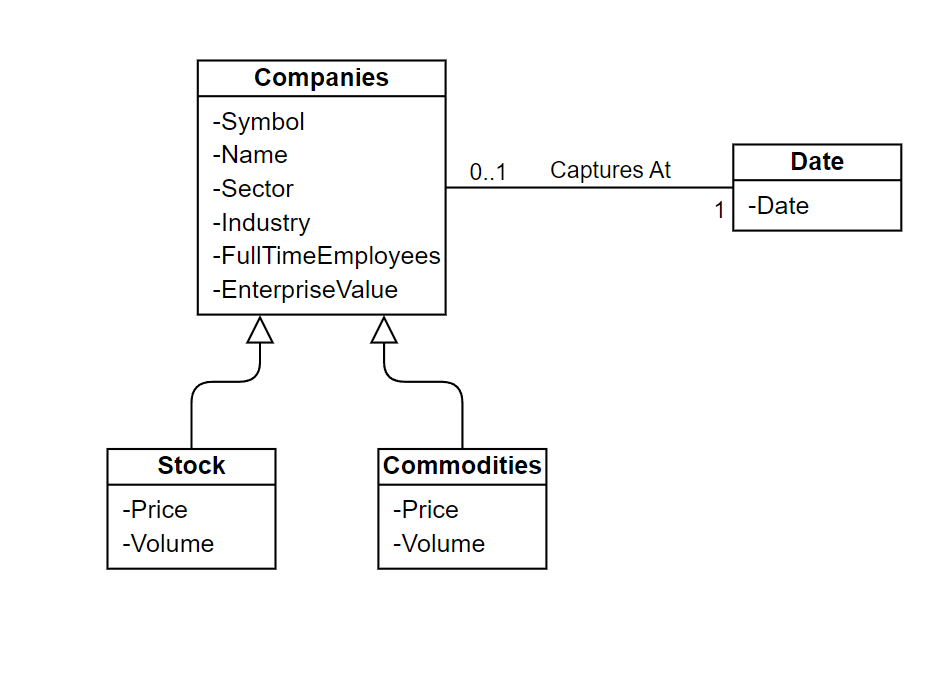}
\caption{Simplified data groups for the analysis and forecasting}
\label{fig: Data Groups}
\end{figure}


As discussed above, in support of the modelling process, data must be collected from various open and commercial sources and then organized in order to (1) create automated ETL routines for the purpose of regular collection of data and (2) develop a data store for supporting the research and model development that is ongoing. Therefore, the ETL process for financial data is an integration and transformation process of data that is being used for extracting data from various sources, transforming it into a homogeneous format, and then loading it into an OLTP DBMS or a DW.


During the data extraction phase, the PHP code sends requests with parameters such as Date, Volume, Open, Close, High, and Low to the financial modelling preparation API. After that, the extracted data is transformed by filtering redundant or unwanted data and converting data types to match the target database's format. 


\begin{table}[ht]
    \centering
    \caption{ETL Load Plan for Data Collection and Processing}
    \begin{tabular}{|l|p{4cm}|}
    \hline
    \textbf{Field} & \textbf{Description} \\
    \hline
    Name & ETL for Data Collection and Processing \\
    \hline
    Description (Optional) & Load plan for collecting data through x, storing it in MySQL, and processing it for analysis using Github and Ionos. \\
    \hline
    Load Plan Type (Optional) & Source Extract and Load (SDE, SIL and PLP) \\
    \hline
    Source Instances (Optional) & Select MySQL as the source instance for data extraction. \\
    \hline
    \end{tabular}
    \label{tab:etl_load_plan}
\end{table}

In the final stage, the data that has been transformed is then loaded into the MySQL database through \cite{ionos.com}\cite{wong2023short}. To maintain database accuracy and keep it up to date, we programmed cron jobs that run every morning for this task.

\subsection{Commodities-ETL Process Workflow Description}

The PHP script retrieves financial data from an external API and stores it in a MySQL OLTP database. The script is designed to connect to the database, retrieve the data from the API, and insert data into the appropriate DBMS tables. Specifically, the data being retrieved consists of stock prices for the Dow Jones, S\&P, NASDAQ, Treasury bonds, gold prices, and crude oil prices between January 1, 2019, and December 31, 2022.

The script defines an array of URLs to retrieve the data from the external API, each corresponding to a specific financial instrument. Then, it loops over this array to retrieve the data using URLs. In addition to the retrieved data from the API, the script defines an array of dates and retrieves additional data for the specified date range.

However, one potential issue with the script is that it does not handle missing data or errors that may occur during the retrieval and insertion of data. It is also unclear how the script fills in missing data, which could lead to incomplete or inaccurate data being stored in the MySQL database.

The script is divided into two parts: (1) The first part retrieves data using an API that provides data every 15 minutes. (2) The second part of the script appears to run to fill in gaps for missing data, although the details of how this is done are unclear.

The script is a basic implementation for retrieving and storing financial data. Still, it could benefit from additional error handling and better documentation on how missing data is being handled and filled in (future work).

\subsection{Commodities-ETL Process code samples}

\begin{lstlisting}[language=PHP, label=Creating the connection]
$mysqli=mysqli_connect(
    <databse_hosting_name>.hosting-data.io,
    <database_user>, <password>, <databsae_name>);

if(mysqli_connect_errno()) {
    tabprintf("Connect failed: %s\n'',
    mysqli_connect_error());
    exit();
}
\end{lstlisting}

\textbf{Creating tables:}
The script first creates tables in the database for each company in the \$companies array. The
table name is constructed by concatenating the company name with the ``-STOCKS" suffix. The table has columns for date, open, high, low, close, and volume and the date column is set as the primary key.

\begin{lstlisting}[language=SQL, label=Creating tables]
foreach($companies as $com){
    $sql="CREATE TABLE IF NOT EXISTS `".$com."-STOCKS`"."
        (`Date` TIMESTAMP,
         `Open` Decimal(8,2),
         `High` Decimal(8,2),
         `Low` Decimal(8,2),
         `Close` Decimal(8,2),
         `Volume` Decimal(30,0),
         PRIMARY KEY (Date));";
    $mysqli->query($sql); 
\end{lstlisting}

\textbf{Retrieving stock data:}
 Stock market data from the financial modelling preparation API is fetched using curl for each company in the table. The API endpoint URL is constructed by using the the company name and the API key as a query parameter. The API returns data in JSON format and it is then decoded into an array.

\begin{lstlisting}[language=PHP, label=Retreiving data]
$des_arr=array(
                'Date'=> array(), 
                'Open'=> array(), 
                'High'=> array(),
                'Low'=> array(),
                'Close'=> array(),
                'Volume'=> array()); 		
$ch = curl_init(
"https://financialmodelingprep.com/api/v3/
    historical-chart/15min/'' . $com .
    "?&apikey=52666840603622de4f9192bcaaa1ce13"
    );
curl_setopt($ch,CURLOPT_RETURNTRANSFER, true);
curl_setopt($ch,CURLOPT_BINARYTRANSFER, true);
$output = curl_exec($ch);
$stock =json_decode($output,true);
$keys=array_keys($stock);

for($i=0;$i<count($keys);$i++){
    $stock_arr['Date'][$i]=
        strtotime($stock[$i]['date']);
    $stock_arr['Open'[$i]=$stock[$i]['open'];
    $stock_arr['High'[$i]=$stock[$i]['high'];
    $stock_arr['Low'[$i]=$stock[$i]['low'];
    $stock_arr['Close'][$i]=
        $stock[$i]['close'];
    $stock_arr['Volume'[$i]=
        $stock[$i]['volume'];
}
\end{lstlisting}

\textbf{Populating tables:}
To insert the stock-market-related data into the database, the script then loops through the array of stock data, and using that, it constructs an SQL query to insert the data into the corresponding table in the database.

\begin{lstlisting}[language=PHP, label=Populating tables]
for($z=0;$z<count($keys);$z++) {
    $sql="INSERT IGNORE INTO `".$com."-STOCKS`" . "
    (`Date`, `Open`, `High`, `Low`, `Close`,` Volume`) 
    VALUES ('".
    date('Y-m-d H:i:s',$stock_arr['Date'][$z])."',".
    $stock_arr['Open'][$z] .",".
    $stock_arr['High'][$z] .",".
    $stock_arr['Low'][$z] .",".
    $stock_arr['Close'][$z] .",".
    $stock_arr['Volume'][$z] .");";
    
    $mysqli->query($sql);
    }
} 
\end{lstlisting}

\section{Data ETL Process Automation}
\subsection{First implementation}
After two years of experiments, we improved our prototype architecture initially described in \cite{wong2023short}:

\begin{figure}
    \centering
    \includegraphics[width=\columnwidth]{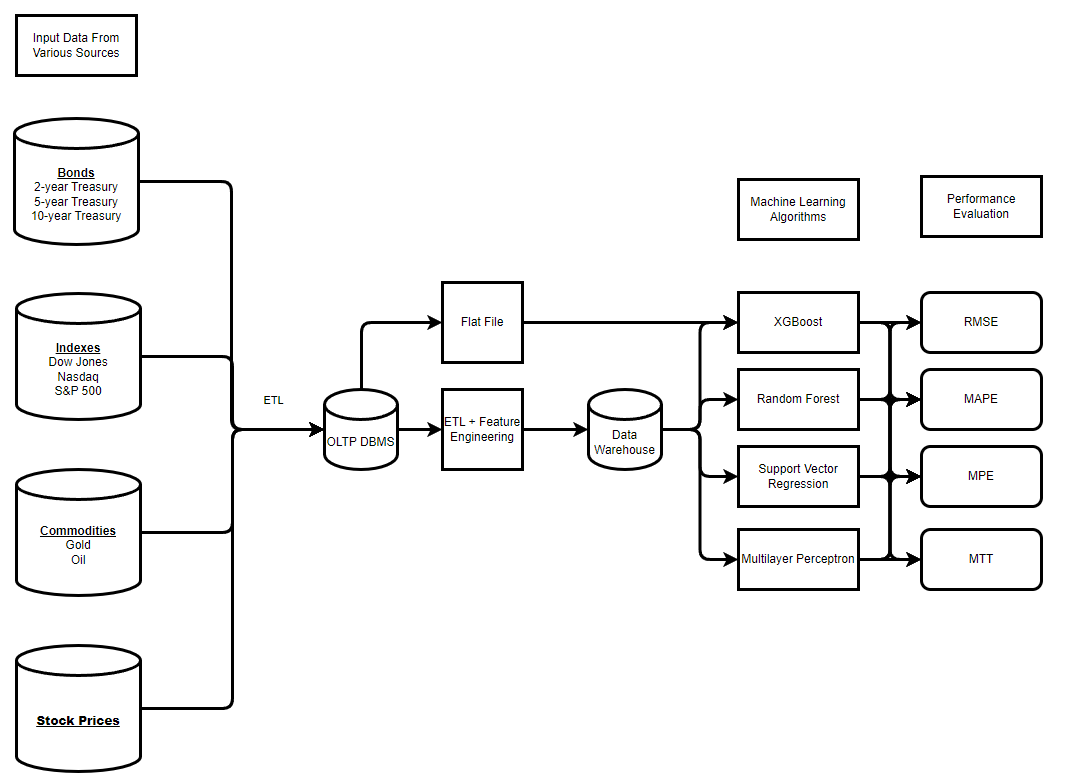}
    \caption{ETL-Automation-Architecture}
    \label{fig: ETL-Architecture}
\end{figure}

 
After the identification of the sources for the data, the data went directly to the feature engineering step. The first ETL process collects and stores the data in MySQL OLTP DBMS. Another ETL process will be used for ETL process data transformation and uploading from OLTP into DW for the ML analysis performance improvement. In the previous iteration, the machine learning algorithms directly used the processed data without utilizing a Data Warehouse. 

We collected and stored data in the OLTP DBMS and prepared data for transformation into DW. Table~\ref{tab:first_data} below presents information about the data processed through the ETL process.
In our first prototype, we collected 5 years of stock data from 2017-08 to 2022-08 in the first two iterations.
\begin{table}[ht]
    \centering
    \caption{Data collected }
    \begin{tabular}{|l|r|r|r|r|r|}
    \hline
    \textbf{Type} & \textbf{Items No.} &  \textbf{Total Records} &
    \textbf{Flat File Size (MB)}\\
    \hline
    Stocks & 359 & 19,910,499 & 1,228.8\\
    \hline
    Commodities & 3 & 359,316 & 26.6\\
    \hline
    Indexes & 3 & 111,900 & 8.24\\
    \hline
   \textbf{ Total }& 365 &  \textbf{20,381,715} &
   \textbf{1,263.64} \\
    \hline    
    \end{tabular}
    \label{tab:first_data}
\end{table} 
This data already occupies 1.26GB with flat files of 
company information. Transformed to a database format, it will occupy a much larger volume. Moreover, 
stocks trading data is considerably larger and 
justifies the qualification of "big data" for our 
project. Our second prototype system makes this 
claim very clearly with the following measurements 
and projection. 

\subsection{Second implementation: data variety}
The second data collection and warehouse 
prototype has the following qualitative and quantitative features. 

Current day stock information (data, price, change\%, change, dayHigh, dayLow, yearHigh, yearLow, mktCap, exchange, open, prevClose, volume).

Historical Stock Information (date, open, high, low, close, adjClose, volume, unadjustedVolume, 
change, changePercentage, vwap, changeOverTime).

Bonds information (country, duration, currency, date, rate) over the past 5 years. 

Current day commodities (date, price, changePercentage, change, dayHigh, dayLow, yearHigh, yearLow, 
mktCap, exchange, open, prevClose, volum
Historical commodities (date, open, high, low, close, adjClose, volume, unadjustedVolume, change, changePercent, vwap, label, changeOverTime). 

Five commodities (copper, crude oil, gold, silver, natural gas). 

Company Statements (date, price, beta, volAvg, mktCap, lastDiv, changes, currency, cik, isin, 
cusip, exchangeFullName, exchange, industry, ceo, sect. 

All the above data types are already 
used effectively by our ML-based 
trading algorithms (our XGBoost 
the model is the most efficient) 
and in the future our ML-based trading 
algorithms will also use sentiment 
analysis, making our systems even 
more comprehensive for their data 
{\em variety}. 

The inclusion of commodities and 
"non-technical" data related to 
trading like company statements, 
makes our approach to warehousing 
and trading a true Big-Data approach: 
{\bf Volume} and {\bf on-the-fly} 
stocks-value prediction coupled 
with true data {\bf variety}. 

\subsection{Second implementation: data volume} 
This system's requirement is to collect 
new data and make them available in the 
data warehouse continuously. 
Its calls to the online data sources 
are limited to 300 API calls per minute 
or 432,000 calls within 24 hours of 
daily operations. From our location in 
B.C. Canada and with the current implementation it takes 7s to execute each collection query, and 5s for each database insertion 
script. As a result, 12s are necessary and sufficient to collect and store the result of 
each data query (as listed above). 
This means our system has time to 
execute 7200 such query-store operations in 24 hours. 

The volume of data can vary with the query and 
its configuration parameters. For example 
up to the past 3 months worth of bond rates, or the past 5 years of a stock's data in a single call. Moreover, queries can support up to 5 symbols (assets) at a time. For a conservative estimate let us 
consider that the 7200 operations per day 
recover one record for every one of 5 symbols, 
which means 36000 records per day. 
We have observed that data occupies about 16.8KB for every 100 records or 172B/record. 
So a day of operation should load and store at the 
very least 5,9MB of data. 
If the system is running 250 days of trading per 
year, it will therefore accumulate a minimum of 
1,44GB for the ongoing year on 5 symbols. 
We also plan to recover and maintain stocks data 
for the last 5 years, a total of 6 times 1,44GB, 
that is 8,64GB for 5 symbols or 1,73GB/symbol of 
continuous storage.


Our system is planned to track at least 500 assets 
so that means continuously updating 6 years' worth 
of data for 500 symbols. As estimated above, this 
represents at the very least 500 times 1,73GB or 
864GB or raw data. 
Database-format storage requirements 
will boost this estimate 
to several terabytes. 

As a result, our approach will be 
able to leverage very large data 
{\em volumes} for on-the-fly 
processing by efficient ML-based 
stocks-value predictions. 
It not only innovates by its many features and 
use of cloud computing resources, but as estimated 
above, also qualifies as a true big-data system 
processing large volume, and variety with (human) real-time velocity. 

\section{Future Works}

\setlength{\fboxrule}{2pt}
This ETL and Automation system prototype can be used alongside the machine learning algorithm for algorithmic trading for a better data pipeline for analysis. As data volumes grow, the scalability and performance of our ETL system become crucial. Future research could explore techniques for handling larger datasets efficiently, potentially utilizing parallel/distributed computing or cloud-based solutions. While we have mentioned the potential use of cloud computing for data processing, future work could delve deeper into cloud integration.

\section{CONCLUSION}
In conclusion, this research has delved into the vital domain of data warehousing and, more specifically, the pivotal role of ETL (Extract, Transform, Load) processes for efficient data management and utilization. Our exploration has been driven by the need to integrate ETL automation with algorithmic trading algorithms seamlessly. This combination holds immense potential in the ever-evolving landscape of financial markets and beyond. This integration promises to provide a reliable way to collect, process, and leverage financial data for algorithmic decision-making.

Automation in ETL processes and data transformation from external sources, uploading in OLTP DBMS and then transforming into DW, is a game-changer, allowing organizations to streamline their data operations, increase efficiency, and enhance data quality. Automating data extraction, cleansing, transformation, and job scheduling is essential for maintaining a competitive edge in today's data-driven landscape.

Moreover, our research has emphasized that the architecture and design of an ETL system are pivotal in data warehousing. Our system exemplifies the importance of a well-structured and versatile design. Its current prototype 
implementations have already 
demonstrated large data volumes, 
data types variety and on-the-fly 
processing coupled with the 
post-processing stocks-price 
prediction systems that have 
already been demonstrated in our 
previous publications.

\section*{ACKNOWLEDGMENTS}
We thank Okanagan College, the OC's Grant in Aid Committee, for the funding and financial support for the  applied student research projects. We are grateful for the funding from Langara College under the Work On Campus (WOC) program, which made some of the research work feasible. In addition, we are equally grateful for the Post-Degree Diploma Program in Data Analytics at Langara College. The program provided two capstone project teams that moved the modelling work forward. We would also like to thank IBM, Oracle, Microsoft, AMAZON, and VMware for supporting us in our educational process, including our research and student projects. The following 
Okanagan College students have 
conducted the second prototype 
implementation, and we thank them for 
their contribution: Jacob Rawlings, Ben Carrier
and Karanmeet Khatra.






\balance

\bibliographystyle{IEEEtran}

\bibliography{2AlgorithmicTr.bib, 2tempBIB.bib, WCCCE23.bib}

\end{document}